# Smart filter aided domain adversarial neural network for fault diagnosis in noisy industrial scenarios

Baorui Dai[1,2], Gaëtan Frusque[2], Tianfu Li[2,3], Qi Li[1,*], Olga Fink[2]

[1]Department of Bridge Engineering, Tongji University, Shanghai 200092, China

[2]Laboratory of Intelligent Maintenance and Operations Systems, EPFL, 1015 Lausanne, Switzerland

[3]School of Mechanical Engineering, Xi'an Jiaotong University, Xi'an 710049, China

[*]Corresponding author. E-mail address: liqi_bridge@tongji.edu.cn (Qi Li).

## Abstract

The application of unsupervised domain adaptation (UDA)-based fault diagnosis methods has shown significant efficacy in industrial settings, facilitating the transfer of operational experience and fault signatures between different operating conditions, different units of a fleet or between simulated and real data. However, in real industrial scenarios, unknown levels and types of noise can amplify the difficulty of domain alignment, thus severely affecting the diagnostic performance of deep learning models. To address this issue, we propose an UDA method called Smart Filter-Aided Domain Adversarial Neural Network (SFDANN) for fault diagnosis in noisy industrial scenarios. The proposed methodology comprises two steps. In the first step, we develop a smart filter that dynamically enforces similarity between the source and target domain data in the time-frequency domain. This is achieved by combining a learnable wavelet packet transform network (LWPT) and a traditional wavelet packet transform module. In the second step, we input the data reconstructed by the smart filter into a domain adversarial neural network (DANN). To learn domain-invariant and discriminative features, the learnable modules of SFDANN are trained in a unified manner with three objectives: time-frequency feature proximity, domain alignment, and fault classification. We validate the effectiveness of the proposed SFDANN method based on two fault diagnosis cases: one involving fault diagnosis of bearings in noisy environments and another involving fault diagnosis of slab tracks in a train-track-bridge coupling vibration system, where the transfer task involves transferring from numerical simulations to field measurements. Results show that compared to other representative state of the art UDA methods, SFDANN exhibits superior performance and remarkable stability.

# 1. Introduction

Intelligent fault diagnosis technology has recently received significant attention and has been widely applied to detect and monitor the health status of mechanical equipment and engineering structures [1-3]. Deep learning algorithms have become a popular choice for intelligent fault diagnosis tasks, with their extensive application being a notable development in this field [4]. Compared to traditional machine learning algorithms that require manual feature engineering, deep learning algorithms can automatically learn meaningful features from raw signals and often achieve better fault diagnosis performance. In particular, deep learning algorithms such as Long Short-term Memory Networks (LSTMs) [5], Generative Adversarial Networks (GANs) [6], and Variational Autoencoders (VAEs) [7] are increasingly gaining prominence in the field of intelligent fault diagnosis. Moreover, in few-shot fault diagnosis scenarios, meta-learning [8] and generative models [9] have recently showcased their effectiveness by learning directly from limited samples and providing data augmentation capabilities, respectively.

Although deep learning-based methods have achieved remarkable success in intelligent fault diagnosis, they still face several challenges that particularly arise from real world limitations [10, 11]. One of the major challenges faced by deep learning-based methods for intelligent fault diagnosis is the difficulty in obtaining sufficient amounts of labeled data. This is particularly due to the fact that faults in safety-critical systems are rare. Insufficient training data can result in overfitting of deep learning models, which limits their applicability in real-world deployments. To address the challenge of limited labeled data availability, one potential approach is to use domain adaptation techniques between a labeled source dataset and an unlabeled target dataset, which may be similar but still exhibit a domain gap due to differences in their distributions [12, 13]. In the context of fault diagnosis, domain adaptation has been employed either between different units of a fleet [14, 15], different operating conditions [16-18], or between simulated and real data [19-22].

Unsupervised domain adaptation (UDA) techniques provide an effective method to address these challenges in fault diagnosis without requiring any labels in target domain [4, 23-25]. The goal of using these techniques is to align the distributions of the source and target datasets, making the features learned from them indistinguishable between the two domains. Leveraging a classifier trained through either the process of feature alignment or based on learned domain-invariant features can be applied to both domains. This approach leads to better fault diagnosis performance in the target domain compared to directly applying models trained on the source domain. In a comprehensive review paper, various UDA methods were compared, showcasing their strong performance in tackling fault diagnosis challenges associated with domain-shifts [10]. Among these methods, those employing adversarial training techniques have shown significant effectiveness in enhancing classification accuracy and have found widespread application across various domains.

While the majority of developed transfer learning approaches has been focusing on closing the domain gap between data captured under real conditions for different operating conditions or different units of a fleet, recently, there has been an increasing emphasis on closing the domain gap between real and simulated data [19, 20, 22, 26]. In the context of simulation-to-real transfer, there are two main directions: one where labels are available in both the source and target domains [26], and another scenario where labels are only available in the source (simulated) domain [22]. Domain transfer from a fully controlled simulated dataset to a real dataset presents several challenges, including unknown class imbalance between different fault types [22] and unknown noise levels. Such challenges can significantly impact the performance of UDA algorithms, and unfortunately, they have not yet been sufficiently addressed. To address the impact of unknown noise, the deep negative correlation multisource domains adaptation networks [27] and marginal denoising autoencoders [28] have been proposed. However, these methods are either tailored for multiple source domain situations or rely on noise-free signals, which limits their broad applicability in real-world industrial settings.

Given that noise can significantly impact the analysis of time-frequency characteristics in signals, one potential approach to mitigate domain gaps is to dynamically modify raw signals in the time-frequency domain during the training of domain adaptation models. This training approach aims to facilitate the learning of shared features between the source and target domains. In the field of signal processing, the recently proposed Denoising Sparse Wavelet Network (DeSpaWN) [29] and the Learnable Wavelet Packet Transform network (LWPT) [30] have shown superior performance in automatically learning meaningful and sparse features from raw signals. This makes them well-suited for unsupervised signal denoising. DeSpaWN primarily emphasizes low-frequency denoising and feature representation, while LWPT evenly prioritizes all frequency ranges. Inspired by these advancements in signal processing, we recently proposed a framework for acoustic signal denoising based on DeSpaWN [31]. By considering vibration signals as denoised variants of acoustic signals, we proposed an effective acoustic signal filtering technique. Additionally, we discovered that by guiding the signal representations, both DeSpaWN and LWPT can approximate one set of signals to another in the time-frequency domain, thereby achieving signal denoising or enhancement. However, it is worth noting that our recently proposed framework requires target labels, which reduces its feasibility in many real-world applications.

In this research, our primary focus is on tackling the challenge of bridging the domain gap between two noisy datasets of high frequency condition monitoring signals, where we only have labels for the source dataset but not for the target dataset. These datasets are characterized by unknown levels and types of noise that are different between the two domains. This significantly affects the alignment between source and target datasets. Specifically, we address the problem of UDA within the context of simulated-to-real domain gap scenarios. In this setup, the source (simulated) dataset would be either noise-free or would only contain controlled and known noise levels, while the target (real) dataset could be severely impacted by noise which is not known a priori. To address this challenge, we propose an unsupervised domain adaptation method called the Smart Filter-Aided Domain Adversarial Neural Network (SFDANN). SFDANN starts with two main components: a learnable wavelet packet transform (LWPT) network and a traditional wavelet packet transform (WPT) module to effectively handle the substantial gap between two domains with different noise levels. By feeding the raw signals from the noisy target domain into the LWPT module, specific coefficients that generate a filtered version of the input signals are produced. Conversely, the raw signals from the source domain are inputted into the WPT module. It gives traditional wavelet coefficients and reconstructs the raw input signals. Then, we introduce a guidance loss that acts on the learnable wavelet coefficients to promote a closer frequency content between the source and (noisy) target domains during the training process. Finally, we input the reconstructed signals of the source and target domains into the latter part of the SFDANN, a typical domain adversarial neural network (DANN) [32], which includes three main parts: a feature extractor, a domain discriminator, and a classifier. The feature extractor extracts features from the source and target domains using a convolutional neural network, the domain discriminator forces the extracted features to lose domain discriminability, and the classifier is trained using the features and labels of the source domain data to classify the aligned source and target domain data. The task of DANN becomes easier due to the similarity in frequency content of both source and target signals, which is achieved with the aid of our proposed smart filter. The proposed SFDANN builds the smart filter, feature extractor, domain discriminator, and classifier in a unified deep neural network framework, allowing the training process of each learnable module to be conducted simultaneously. The main contributions of our research are summarized as follows:

(1) We propose a smart filter that combines the LWPT and WPT techniques with our

developed guidance loss. This smart filter dynamically enforces the similarity between the source and (noisy) target domain signals in the time-frequency domain during the training process. To the best of the authors' knowledge, it is the first time that the LWPT and WPT techniques are used in the context of UDA to denoise signals.

(2) Our proposed UDA method utilizes the smart filter to enhance the performance of the domain adversarial neural network for end-to-end intelligent fault diagnosis. This approach effectively captures aligned features between the source and target domains, especially in noisy industrial scenarios.

(3) In our extensive evaluations, we specifically focus on two main application scenarios that have been rarely studied: (a) UDA between two datasets with different levels and types of noise impact in the source and target domain signals, and (b) UDA between a noise-free simulated dataset and a real-world dataset with an unknown level and type of noise. In our application, we consider the train-track-bridge coupling system as a specific case study.

The remaining content of this paper is structured as follows. In Section 2, we provide the background on the DANN and the LWPT. Section 3 presents our proposed method, SFDANN. In Section 4, we demonstrate the effectiveness of SFDANN in cross-domain fault diagnosis through two case studies. Section 4 also includes ablation studies on SFDANN and explores the specific role of the smart filter. Finally, in Section 5, we present the main conclusions of our work.

## 2. Background

*2.1. Domain adversarial neural network (DANN)*

DANN is a neural network inspired by generative adversarial networks and has found widespread applications in various fields, including fault diagnosis [33] and prognosis [34], speech recognition [35], and sentiment analysis [36]. It comprises a feature extractor ($G_F$), a domain discriminator ($G_D$), and a classifier ($G_C$) as depicted in Fig. 1 [32]. DANN captures transferable features that are invariant to domain changes through a minimax game between $G_D$ and $G_F$. It learns features that are effective for classification by leveraging the relationship between $G_C$ and $G_F$. Consequently, DANN can extract features that are both discriminative for classification tasks and invariant to domain changes, making it well-suited for various cross-domain fault diagnosis applications.

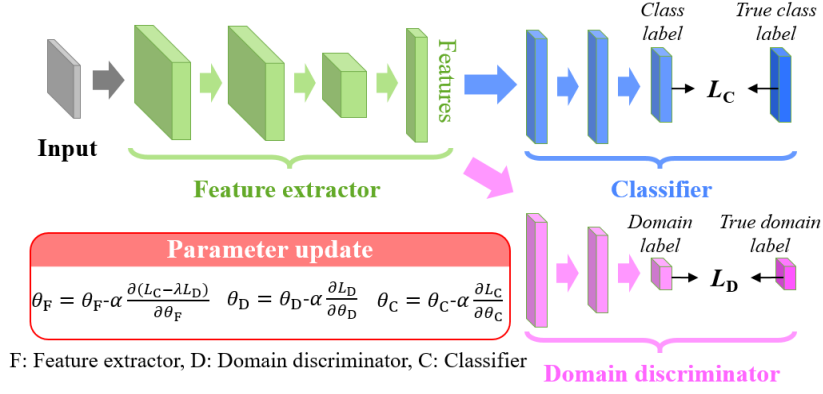

F: Feature extractor, D: Domain discriminator, C: Classifier

Fig. 1. Architecture of classic DANN.

The labeled data from the source domain is defined as $\{x_i^s; y_i^s\}_{i=1}^{N_s}$, and the unlabeled data from the target domain is defined as $\{x_i^t\}_{i=1}^{N_t}$. Here, $x_i^s$, $y_i^s$, and $x_i^t$ represent the *i*-th sample from the source domain, the corresponding label for the *i*-th sample from the source domain, and the *i*-th sample from the target domain, respectively. $N_s$ and $N_t$ represent the number of samples from the source domain and target domain. DANN uses a loss function to minimize classification error and maximize domain discrimination error, thereby improving the classification ability of models and reducing the differences in domain distribution of extracted features. This loss function consists of the classification loss $L_C$, defined by the cross-entropy loss, and the domain discrimination loss $L_D$, defined by the binary cross-entropy loss. It can be expressed as:

$$L(\theta_F, \theta_C, \theta_D) = L_c(\theta_F, \theta_C) - \lambda L_D(\theta_F, \theta_D) \tag{1}$$

$$L_C(\theta_F, \theta_C) = -\frac{1}{N_s}\sum_{i=1}^{N_s}\sum_{c=0}^{C-1} I[y_i^s = c]\log\big(G_C(G_F(x_i^s; \theta_F); \theta_C)\big) \tag{2}$$

$$L_D(\theta_F, \theta_D) = -\frac{1}{N_s}\sum_{i=1}^{N_s} \log\big(G_D(G_F(x_i^s; \theta_F); \theta_D)\big) - \frac{1}{N_t}\sum_{i=1}^{N_t} \log\big(1 - G_D(G_F(x_i^t; \theta_F); \theta_D)\big) \tag{3}$$

where $\theta_F$, $\theta_C$, and $\theta_D$ are the parameters of $G_F$, $G_C$, and $G_D$, respectively; $\lambda$ is a trade-off parameter; $I[\cdot]$ is a symbol function that takes a value of 1 if the true label of $x_i^s$ is $c$, and 0 otherwise; and $C$ is the number of class labels.

## 2.2. Learnable wavelet packet transform network (LWPT)

LWPT is a recently proposed deep learning framework that draws inspiration from WPT and aims to automatically learn meaningful and sparse representations of raw signals [30]. Fig. 2 depicts the cascade algorithm associated with WPT, which serves as the fundamental architecture of LWPT. This algorithm decomposes the input signal into detail and approximation coefficients by applying a low-pass and a high-pass filter, followed by a sub-sampling step. In a recursive manner, the detail and approximation coefficients of the previous layer are decomposed using a similar process. The detail and approximation coefficients of the final decomposition layer form the time-frequency representation of the input signal. By using the inverse WPT, the input signal can be perfectly reconstructed from the obtained representation.

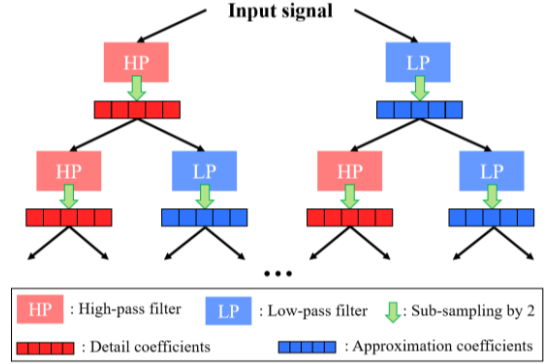

Fig. 2. Architecture of the cascade algorithm related to the WPT.

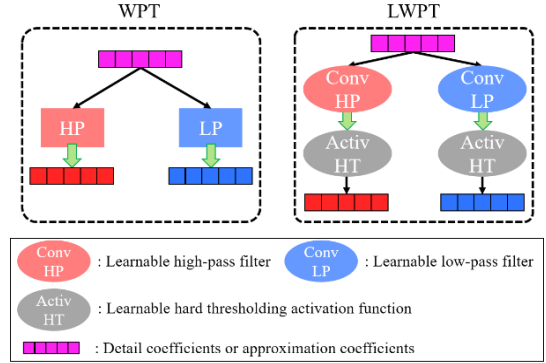

Fig. 3. Encoding blocks of the WPT and the LWPT.

LWPT adopts an encoder-decoder architecture that sequentially applies learnable signal decomposition and learnable signal reconstruction. It incorporates a fully learnable version of the cascade algorithm, enabling the learning of the kernel shared by both filters at each layer. The resulting detail and approximation coefficients from each layer are then passed through learnable hard thresholding (HT) activation functions originally proposed for DeSpaWN [29]. Fig. 3 illustrates the encoding blocks of WPT and LWPT, with elements shared with Fig. 2. The HT activation function is a combination of two sigmoid functions with opposite characteristics, and it is mathematically expressed as follows:

$$\text{HT}(x) = x \left[ \frac{1}{1+\exp(10(x+b))} + \frac{1}{1+\exp(-10(x-b))} \right] \tag{4}$$

where $b$ is the learnable bias acting as the thresholds on both sides of the origin.

The decoding blocks of LWPT exhibit a reverse architecture compared to the encoding blocks but without the HT activation functions. In each layer, the learnable HT activation functions operate independently to automatically denoise the wavelet coefficients in the encoding blocks. However, due to the denoising effect of HT activation function, achieving perfect signal reconstruction with the decoding part of LWPT is not feasible anymore.

## 3. Proposed method

*3.1. Overall framework of the proposed method*

When the signals are significantly affected by high levels of noise, current domain adaptation methods may face challenges in effectively learning the shared features between the source and target domains, leading to misclassification of a significant number of samples in the target domain. In this research, we focus on high frequency condition monitoring signals. To facilitate the learning of domain-invariant and discriminative features, we propose dynamically enforcing the similarity between the source and target domain signals in the time-frequency domain during the training process while learning to filter the noisy target domain signals. The overall framework of our proposed SFDANN is depicted in Fig. 4, comprising four modules: a smart filter, a feature extractor, a domain discriminator, and a classifier. To clarify the operation mechanism of the SFDANN, the architecture in Fig. 4 is divided into two parts: Part A and Part B. Part A comprises the data input and the smart filter, which can filter signals without requiring neither labels nor ground truth noise-free signals. Part A inputs the source domain data into the WPT module and the (noisy) target domain data into the LWPT module. We also evaluate the opposite data input strategy, where we input the (noisy) target domain signals into the WPT and the source domain data into the LWPT. It is important to note that Fig. 4 displays only the first of the data input strategies. The optimal data input strategy will be further discussed in the ablation study section based on the evaluation of the obtained results. Part B involves the typical DANN architecture. During the training stage, the data from the source and target domains undergo filtering in Part A before being passed to Part B. Additionally, loss functions in Part B affect the parameter learning in Part A through error backpropagation.

As depicted in Fig. 4, firstly, raw signals from the source domain are input into the WPT module's encoder for signal decomposition, while the raw signals from the target domain are input into the LWPT module's encoder for signal decomposition. This process generates traditional wavelet coefficients and learnable wavelet coefficients, respectively. Next, the traditional wavelet coefficients and learnable wavelet coefficients are reconstructed into time-domain signals by the decoders of the WPT module and the LWPT modules. Guided by the guidance loss, the features of the learnable wavelet coefficients dynamically converge towards those of the traditional wavelet coefficients during the learning process. This convergence enforces the similarity between the source and target domain signals in the time-frequency domain while filtering noisy signals.

After being processed by the smart filter in Part A, the time-domain signals from the source and target domains are fed into the Part B of SFDANN. Part B consists of three typical components found in a DANN: feature extractor, domain discriminator, and classifier. The feature extractor utilizes a convolutional neural network to extract features from the source and target domains. The domain discriminator, guided by the domain discrimination loss, encourages the extracted features to lose domain distinction and become indistinguishable between the two domains. Meanwhile, the classifier is trained to minimize the misclassification loss associated with the source domain data. Moreover, the learning process of the smart filter is also guided by the domain discrimination loss and the classification loss, ensuring that the filter adjusts the time-frequency characteristics to facilitate the learning of domain-invariant and discriminative features.

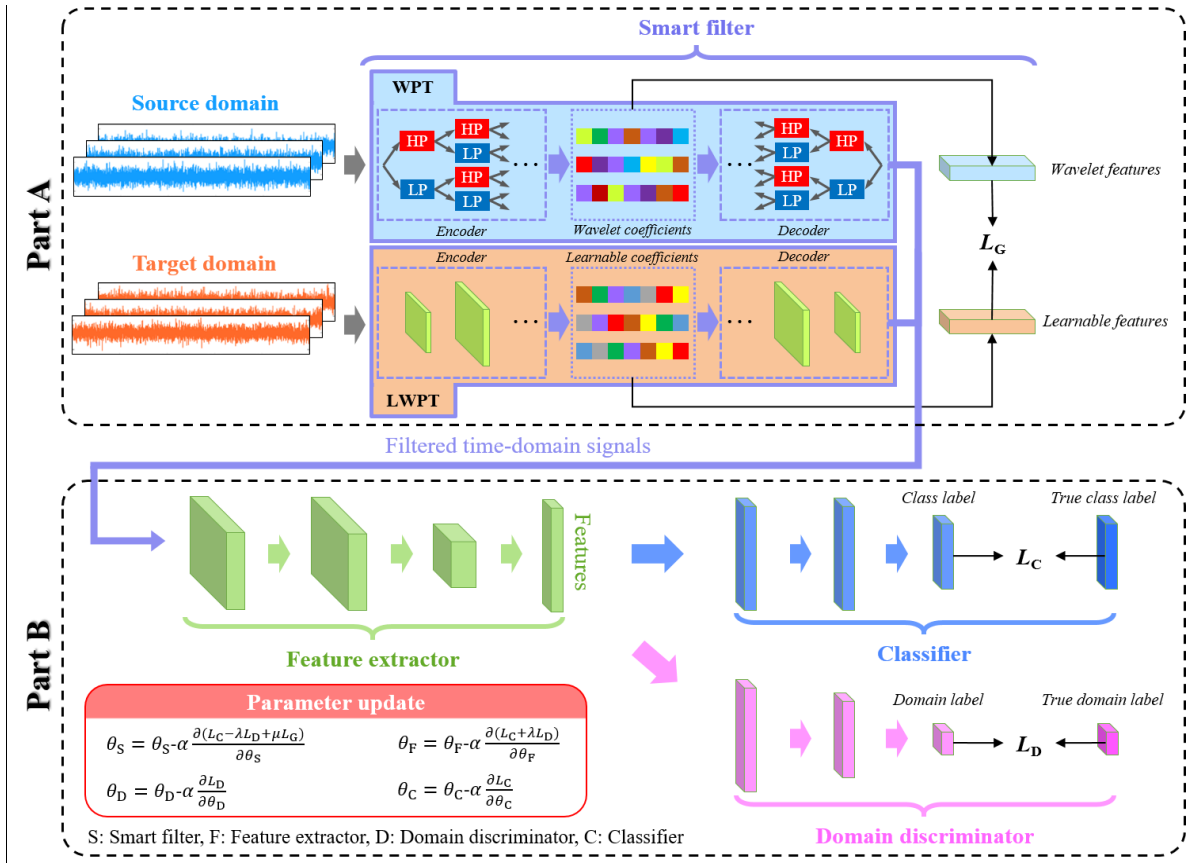

Fig. 4. Architecture of the proposed SFDANN.

### 3.2. Smart filter

The smart filter $G_s$ consists of two modules: $G_{s-WPT}$ (WPT module) and $G_{s-LWPT}$ (LWPT module). Two distinct data input strategies are viable for the smart filter, the optimal of which will be discussed in the ablation study section. For simplicity, we will explain the functioning of the smart filter using an example where data from the source domain is fed into $G_{s-WPT}$, and data from the target domain is fed into $G_{s-LWPT}$, consistent with depicted in Fig. 4.

First, $x_i^s$ is decomposed into traditional wavelet coefficients $\left\{\left\{c_{i,j,k}^s\right\}_{k=1}^K\right\}_{j=1}^{2^L}$ using the encoder in $G_{s-WPT}$, which follows an *L*-layer signal decomposition framework. Similarly, $x_i^t$ is decomposed into learnable wavelet coefficients $\left\{\left\{c_{i,j,k}^t\right\}_{k=1}^K\right\}_{j=1}^{2^L}$ using the encoder in $G_{s-LWPT}$, which also has an *L*-layer signal decomposition framework. Here, *K* is the number of wavelet coefficients. The determination of the number of decomposition layers *L* is influenced by the sampling frequency $F_s$ and the frequency bandwidth $F_r$, which are important for fault diagnosis of the input signal, as stated in [29, 30]. The determination of $F_r$ is dependent on the inherent characteristics of the fault diagnosis object and will be further discussed in the case studies section. It is essential to highlight that when increasing the number of signal decomposition layers *L*, it directly adds complexity to the convolutional neural network. While this can result in finer decomposition of source and target domain signals, encouraging similarity in these smaller frequency bands, it is crucial to note that these fine-grained frequency bands do not possess specific physical significance for fault diagnosis. Consequently, they do not contribute to improving the smart filter's performance. On the contrary, the augmented learnable parameters can elevate the learning difficulty due to potential overfitting effect, especially when the dataset has limited samples.

In real industrial scenarios, significant differences in time-frequency features between the source and target domain signals often arise due to factors such as noise interference. These differences typically stem from the distinct environments of each domain. Consequently, it is desirable to achieve overall similarity in the time-frequency characteristics of source and target domain samples. The expectation of the features of all samples from the source domain provides an overall representation of the data characteristics in that domain. However, during training, deep learning models typically process data in batches rather than the entire dataset at once. A batch size, which is smaller than the total number of samples, is determined, and a random batch of samples is selected for each training iteration. To ensure that the expectation of the features in a batch approximates that of the entire dataset, the ratio of samples corresponding to each class in the batch should be roughly equivalent to that in the whole dataset. According to the law of large numbers (specifically, Bernoulli's law of large numbers), it is important to ensure that the batch size is not too small. Further details regarding the appropriate batch size will be presented in the case studies section.

Assuming that the batch size of SFDANN is $B$, we define a guidance loss to encourage the expectation of the learnable wavelet coefficients $\left\{\left\{\left\{c_{i,j,k}^{\text{t}}\right\}_{k=1}^{K}\right\}_{j=1}^{2^L}\right\}_{i=1}^{B}$ for a batch of target domain samples $\{x_i^{\text{t}}\}_{i=1}^{B}$ to be close to the expectation of the traditional wavelet coefficients $\left\{\left\{\left\{c_{i,j,k}^{\text{s}}\right\}_{k=1}^{K}\right\}_{j=1}^{2^L}\right\}_{i=1}^{B}$ for a batch of source domain samples $\{x_i^{\text{s}}\}_{i=1}^{B}$. The guidance loss is expressed as follows:

$$L_{\text{G}}(\theta_{\text{S}}) = \frac{1}{2^L}\sum_{j=1}^{2^L}\left(\text{E}\left[\left\{\left\{c_{i,j,k}^{\text{s}}\right\}_{k=1}^{K}\right\}_{i=1}^{B}\right] - \text{E}\left[\left\{\left\{c_{i,j,k}^{\text{t}}\right\}_{k=1}^{K}\right\}_{i=1}^{B}\right]\right)^2 \tag{5}$$

$$\text{E}\left[\left\{\left\{c_{i,j,k}^{\text{s}}\right\}_{k=1}^{K}\right\}_{i=1}^{B}\right] = \frac{1}{B}\sum_{i=1}^{B}\frac{1}{K}\sum_{k=1}^{K}\left|c_{i,j,k}^{\text{s}}\right| \tag{6}$$

$$\text{E}\left[\left\{\left\{c_{i,j,k}^{\text{t}}\right\}_{k=1}^{K}\right\}_{i=1}^{B}\right] = \frac{1}{B}\sum_{i=1}^{B}\frac{1}{K}\sum_{k=1}^{K}\left|c_{i,j,k}^{\text{t}}\right| \tag{7}$$

where $\theta_{\text{S}}$ represents the parameters of the smart filter $\text{G}_{\text{S}}$; $k$ is the wavelet coefficient number, which is related to time.

While being guided by the guidance loss $L_{\text{G}}(\theta_{\text{S}})$ and obtaining a similar signal content representation between source and target, the traditional wavelet coefficients $\left\{\left\{\left\{c_{i,j,k}^{\text{s}}\right\}_{k=1}^{K}\right\}_{j=1}^{2^L}\right\}_{i=1}^{B}$ and learnable wavelet coefficients $\left\{\left\{\left\{c_{i,j,k}^{\text{t}}\right\}_{k=1}^{K}\right\}_{j=1}^{2^L}\right\}_{i=1}^{B}$ are input into the separate decoders with $L$-layer signal reconstruction frameworks in $\text{G}_{\text{S-WPT}}$ and $\text{G}_{\text{S-LWPT}}$, respectively, to generate reconstructed signals $\{\hat{x}_i^{\text{s}}\}_{i=1}^{B}$ and $\{\hat{x}_i^{\text{t}}\}_{i=1}^{B}$ in the source and target domains. Then, $\{\hat{x}_i^{\text{s}}; y_i^{\text{s}}\}_{i=1}^{B}$ and $\{\hat{x}_i^{\text{t}}\}_{i=1}^{B}$ are input into the subsequent DANN framework.

*3.3. Objective function*

To facilitate the learning of domain-invariant and discriminative features, our SFDANN is trained using three loss functions: the guidance loss, classification loss, and domain alignment loss. These loss functions directly and individually impact the smart filter, classifier, and domain discriminator components of SFDANN, as depicted in Fig. 4.

*3.3.1. Guidance loss*

The guidance loss serves the purpose of dynamically aligning the time-frequency features of the source and target domains, facilitating the learning of shared features between them. The expression for the guidance loss is shown in Eq. (5).

*3.3.2. Classification loss*

The classification loss is calculated using cross entropy loss, which measures the discrepancy between the predicted label and the true label. This loss function guides the classifier to improve its classification performance and make accurate predictions. The expression for the classification loss is provided in Eq. (2). However, in our proposed SFDANN, the input $x_i^s$ in Eq. (2) is replaced by $\hat{x}_i^s$, which corresponds to $G_S(x_i^s; \theta_S)$. As a result, $L_C(\theta_F, \theta_C)$ in Eq. (2) is modified to $L_C(\theta_S, \theta_F, \theta_C)$.

*3.3.3. Domain alignment loss*

The domain alignment loss is calculated using binary cross-entropy loss, which quantifies the discrepancy between the source and target domain labels. Its objective is to ensure that the features extracted by the feature extractor are distinguishable by the domain discriminator. The expression for the domain alignment loss is presented in Eq. (3). However, in our proposed SFDANN, the terms $x_i^s$ and $x_i^t$ in Eq. (3) are substituted by $\hat{x}_i^s$ and $\hat{x}_i^t$, respectively, denoting the outputs of $G_S(x_i^s; \theta_S)$ and $G_S(x_i^t; \theta_S)$. As a result, $L_D(\theta_F, \theta_D)$ in Eq. (3) is modified to $L_D(\theta_S, \theta_F, \theta_D)$.

*3.3.4. Overall objective function*

The overall objective function of SFDANN is a combination of the three loss functions and can be expressed as follows:

$$L(\theta_S, \theta_F, \theta_D, \theta_C) = L_C(\theta_S, \theta_F, \theta_C) - \lambda L_D(\theta_S, \theta_F, \theta_D) + \mu L_G(\theta_S) \tag{8}$$

where $\lambda$ and $\mu$ are the trade-off parameters.

By assigning higher importance to $L_G$ than $L_D$ in the initial stages of training, we can facilitate the alignment of time-frequency features between the source and target domains and create favorable conditions for extracting shared features from them. Following the approach proposed by [37], we introduce a weight expression for $L_D$, denoted as $\lambda$, which increases from 0 to 1 as the training progresses:

$$\lambda = \frac{2}{1+e^{-10p}} - 1 \tag{9}$$

where $p$ is the ratio of the current epochs to the total number of epochs in the training procedure.

## 4. Case studies

*4.1. Introduction of case studies*

We validate the effectiveness of the proposed SFDANN for fault diagnosis in industrial scenarios with unknown types and levels of noise through two cases. The first case study focuses on fault diagnosis of bearings, where the source and target domain data are constructed with a modified version of CWRU dataset [38] to match the case of significantly different levels of noise interference. The second case study involves the recognition of the health states of slab track in a train-track-bridge coupling system, where the source data are obtained from numerical simulation and target domain data comprise field measurements. In both cases, the data utilized consists of acceleration signals.

To evaluate the performance of SFDANN, we compare it with five other UDA methods: joint adaptation networks (JAN) [39], multi kernels maximum mean discrepancy (MK-MMD) [40], correlation alignment (CORAL) [41], DANN [32], and conditional DANN (CDANN) [42]. To ensure fair and effective comparisons, we use a unified testing framework and consistent parameter settings for all methods, following the structure outlined in the review paper by [10]. A four-layer CNN serves as the feature extractor for all methods, and we also use this feature extractor in combination with a classifier as a baseline for comparison with UDA methods. In the case of the SFDANN method, unless stated otherwise, the source domain data is input into the WPT module and target domain data into the LWPT module, as depicted in Fig. 4. Since the classes in both case studies are balanced, the performance evaluation metric used in this paper is the overall accuracy of classification, calculated as the number of accurately classified samples divided by the total number of samples. To minimize result variability, we perform calculations with five random seeds and report the average and standard deviation of the classification results.

*4.1.1. Case Study 1: bearing fault diagnosis*

*(1) Data description*

The CWRU dataset has been one of the most commonly used open-source datasets in UDA research. A diagram illustrating the experimental setup is provided in Fig. 5 [38]. In this research, we use the 12 kHz drive-end bearing data, with detailed parameters provided in Table 1. The bearing data consists of four different operating conditions, each corresponding to a different combination of load and rotational speed parameters. The dataset contains 10 different health conditions of the bearing, including one healthy state and three different fault types (inner ring faults (IF), outer ring faults (OF) and ball faults (BF)) with three different fault sizes each.

However, unlike previous studies, our case considers the impact of environmental noise on the target domain data, which is considered more realistic for industrial applications. In practical industrial settings, various levels of noise interference are unavoidable, and UDA techniques always involve generalizing data to unknown levels of noise. To simulate real-world scenarios, we add Gaussian white noise to target domain data, resulting in noisy signals with signal-to-noise ratios (SNRs) of 0 or -5. The SNR is defined as follows:

$$\text{SNR}_{\text{dB}} = 10 \log_{10} \frac{P_\text{s}}{P_\text{n}} \tag{10}$$

where $P_\text{s}$ and $P_\text{n}$ are the power of the signal and noise, respectively.

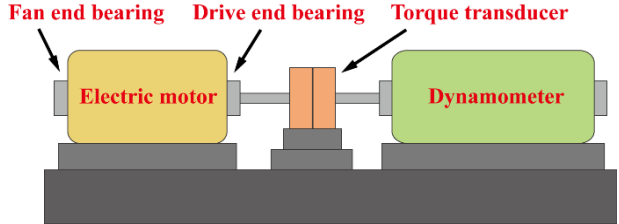

Fig. 5. Diagram of the test rig for collecting CWRU data.

Table 1. Parameters for the CWRU dataset.

| Operation condition | Rotational speed (RPM) | Class label (Fault type-severity in mm) |
|---|---|---|
| 0 | 1797 | 0(NA), 1(IF-7), 2(BF-7), 3(OF-7), 4(IF-14), 5(BF-14), 6(OF-14), 7(IF-21), 8(BF-21), 9(OF-21) |
| 1 | 1772 | |
| 2 | 1750 | |
| 3 | 1730 | |

*(2) Parameter settings*

In this study, the Z-score normalization method is used to normalize the data. The normalized signal is then segmented into samples using a sliding window with a window length of 2048, which coincides with the approach and parameters adopted by [43] and [44]. The segmentation process does not involve any overlap between the samples. Consequently, each fault state contains approximately 60 samples, while the healthy state contains approximately 120 samples under operation condition 0 and 240 samples under other operation conditions. For both the source and target domains, 80% of the total samples are used for training, and 20% are reserved for testing. The cross-domain task of 0→1 is denoted as T01, indicating that the model is trained using labeled training data from operation condition 0 and unlabeled training data from operation condition 1, and subsequently tested using testing data from operation condition 1.

The model training process is conducted over a total of 300 epochs. The initial learning rate is set to 0.001, which is reduced by a factor of 0.1 at epochs 150 and 250 to facilitate learning rate decay. To ensure the inclusion of meaningful frequency bandwidth for bearing fault diagnosis, the smart filter is configured with five signal decomposition layers, resulting in a frequency resolution of $12000/2/2^5 = 187.5\text{Hz}$. The batch size is set to 64. The trade-off parameter $\lambda$ is determined using Eq. (9), while the value of $\mu$ is adjusted based on the severity of the noise. Specifically, $\mu$ is set to 1 when no Gaussian white noise is present, 2 at a SNR of 0, and 10 at a SNR of -5.

*4.1.2. Case Study 2: State recognition of slab tracks*

Slab track has been increasingly built in urban rail transit and high-speed railways. The mortar layer, which connects the slab track and the foundation, is susceptible to degradation due to train dynamic loads, temperature fluctuations, and other factors. Recognizing the state of the slab track is crucial for ensuring the safe operation of trains. However, acquiring an adequate amount of labeled data from field measurements poses challenges. Moreover, measurements are usually taken under noisy environments. Therefore, conducting numerical simulation and then applying the UDA methods to transfer from labeled and noise-free numerical simulation data to unlabeled and noisy measurement data is very important.

*(1) Field measurements*

Since the primary cause of slab track deterioration is the degradation of the mortar layer, which alters the support conditions of the slab track, we have chosen three different support conditions of slab track from a railway test line to represent three different deterioration states, as depicted in Fig. 6. The connections between the three track slabs and the foundation are supported by mortar, rubber, and discrete spring, respectively, representing the healthy state, moderate deterioration state, and severe deterioration state.

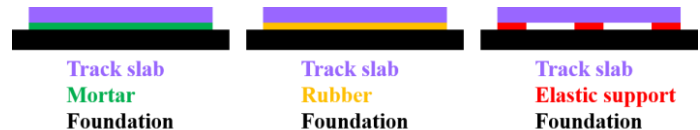

Fig. 6. Three types of slab track with different support conditions.

The railway test line is installed on simply supported girder bridges. A six-vehicle metro train with a total length of 140 meters operates on the railway test line. Acceleration signals are collected from acceleration sensors installed on the three types of slab track, as depicted in Fig. 7. The signals are sampled at a frequency of 20 kHz. The train operates at speeds of 20, 40, 60, and 80 km/h, and six passes are conducted at each speed.

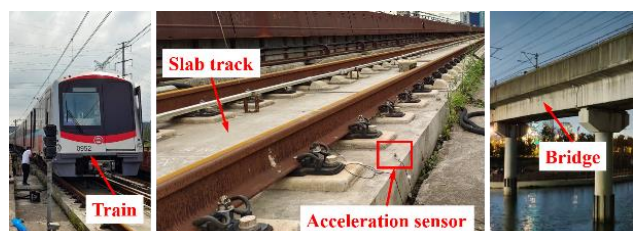

Fig. 7. Placement of acceleration sensors.

*(2) Numerical simulation*

To obtain the numerical simulation signals of slab track vibration, a coupling vibration numerical model is established based on the mechanical parameters of the train, track, and bridge obtained from the field experiment. The model development is divided as two steps to reduce computational costs. Firstly, the numerical model of the train, steel rail, and bridge is established using ANSYS software with the beam elements, as depicted in Fig. 8. To generate track irregularity, we use the irregularity spectrum from ISO 3095-2013 [45] and then obtain the wheel-rail force time-history data using the modal superposition method [46] when the train travels at the four different speeds.

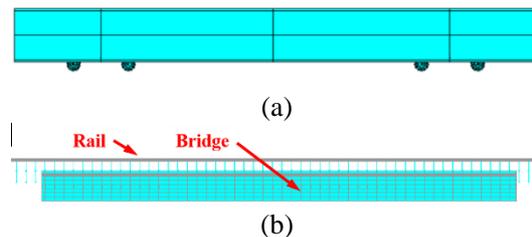

Fig. 8. Numerical models with beam element: (a) Single vehicle of metro train; (b) Rail and bridge.

Secondly, a refined numerical model of the steel rail, slab track, and bridge is developed using the ABAQUS software. Beam elements are used to model the steel rails, while solid elements are used to model the slab track and bridge structures, as illustrated in Fig. 9. The wheel-rail force time-history data obtained previously is loaded onto the numerical model in ABAQUS as a moving load. The simulated acceleration signals of the slab track are then calculated corresponding to the placement position of the acceleration sensors in the field measurement.

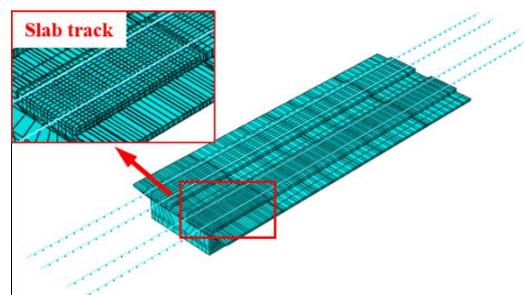

Fig. 9. Refined numerical models of rail, slab track and bridge.

*(3) Parameter Settings*

In the time domain, the acceleration signal of the slab track exhibits six cycles of waveforms due to the continuous excitation caused by the six vehicles of the metro train. To enhance the dataset, each measured or simulated sample is divided into six sub-samples based on the time when each vehicle passes through the slab track. Consequently, for both measured and simulated data, there are a total of 36 sub-samples for each track state at each speed.

In the frequency domain, the computational cost significantly increases when calculating high-frequency vibrations. Therefore, we set the maximum frequency of the numerical simulation to 800Hz. To ensure consistency with the frequency range of the numerical simulation signal, we apply a bandpass filter to the measured signal and retain the frequency components below 800Hz.

For data normalization, we use the Z-score normalization method to globally normalize the data, using the numerical simulation as the source domain and the field measurement as the target domain. In both domains, 75% of the total samples are allocated for the training set, while the remaining 25% are assigned to the testing set. To approximate more realistic scenarios, sub-samples from the same train ride are either all included in the training dataset or all included in the testing dataset.

In this case study, we use healthy and faulty acceleration signals generated from numerical simulations as input source domain data, while healthy and faulty signals collected from acceleration sensors during field experiments serve as input target domain data. The resulting output of the SFDANN provides classification labels. The model training parameters are the same as those used in Case Study 1. The smart filter includes six signal decomposition layers, which corresponds to a frequency resolution of $800/2^6 = 12.5$Hz. The batch size is set to 27, ensuring that it can be evenly divided by $36 \times 3 \times 75\% = 81$ training samples at each speed. The trade-off parameter $\lambda$ is set according to Eq. (9), and $\mu$ is set to 5.

### 4.2. Results and analysis

#### 4.2.1. Diagnosis results

*(1) Diagnosis results of Case Study 1 (Bearing dataset)*

Diagnosis results in the Case Study 1 are obtained considering noise added in the target domain data with SNRs of 0 and -5. Given the extensive diagnosis results obtained through experiments conducted with various noise levels and based on different UDA methods, we selectively focus on a subset of transfer tasks. Specifically, we present transfer diagnosis results from operating condition 0 to other operating conditions, as well as from operating condition 1 to other operating conditions. This subset constitutes half of all transfer tasks and provides a suitable basis for comparing the performance of different UDA methods. Tables 2 showcases these results. The findings in Tables 2 clearly demonstrate the superior fault diagnosis accuracy achieved by the SFDANN method proposed in this paper. When using the original noise-free CWRU dataset, all UDA methods attain fault diagnosis accuracy close to 100%, indicating the effective resolution of domain transfer challenges through conventional UDA methods in the original CWRU dataset. However, the introduction of noise interference into the target domain dataset adversely impacts the performance of all fault diagnosis methods. At an SNR of 0, the SFDANN method exhibits an average diagnostic accuracy of 96.3%, surpassing other fault diagnosis methods by 2.0%-50.8%. At the SNR of -5, the superiority of the SFDANN method over other fault diagnosis methods becomes more pronounced. Despite the average diagnostic accuracy of the SFDANN method decreasing to 83.9%, it continues to outperform other fault diagnosis methods by 10.7%-52.9%. These results underscore the exceptional diagnostic performance of the SFDANN method in noisy environments.

Table 2. Bearing fault diagnosis accuracy considering various transfer tasks and noise levels.

| Noise level | Task | CNN | CORAL | MK-MMD | JAN | DANN | CDANN | SFDANN |
|---|---|---|---|---|---|---|---|---|
| No noise | T01 | 99.4%±1.3% | **100.0%±0** | **100.0%±0** | 99.9%±0.3% | 99.9%±0.3% | **100.0%±0** | 99.9%±0.3% |
| | T02 | 99.6%±0.3% | **99.9%±0.3%** | 99.8%±0.3% | 99.7%±0.3% | 99.8%±0.3% | 99.6%±0.3% | 99.7%±0.3% |
| | T03 | 86.5%±2.5% | **100.0%±0** | 99.9%±0.3% | 99.7%±0.4% | 99.8%±0.3% | 99.8%±0.3% | 99.7%±0.3% |
| | T10 | 98.9%±1.1% | 99.7%±0.4% | 99.8%±0.3% | 99.5%±0.4% | 99.8%±0.3% | 99.5%±0.4% | **99.9%±0.3%** |
| | T12 | 99.5%±0.3% | 99.7%±0.3% | 99.7%±0.3% | 99.7%±0.3% | **99.9%±0.3%** | **99.9%±0.3%** | **99.9%±0.3%** |
| | T13 | 97.1%±1.2% | 99.8%±0.3% | **100.0%±0** | 99.9%±0.2% | 99.8%±0.2% | **100.0%±0** | 99.9%±0.3% |
| | Average | 96.8%±1.1% | **99.9%±0.2%** | **99.9%±0.2%** | 99.7%±0.3% | 99.8%±0.3% | 99.8%±0.2% | 99.8%±0.3% |
| SNR=0 | T01 | 43.1%±12.7% | 31.1%±21.0% | 93.4%±2.3% | 93.8%±1.2% | 95.5%±1.8% | 93.6%±2.4% | **96.1%±1.0%** |
| | T02 | 43.1%±8.1% | 56.1%±24.1% | 93.7%±3.3% | 94.8%±1.7% | 95.1%±1.6% | 93.1%±1.8% | **97.1%±1.0%** |
| | T03 | 30.3%±5.2% | 22.7%±5.1% | 75.8%±2.8% | 90.5%±3.0% | 91.9%±4.5% | 88.3%±4.1% | **96.2%±1.3%** |
| | T10 | 50.6%±10.7% | 58.6%±1.5% | 82.9%±2.9% | 77.9%±4.9% | 93.9%±2.3% | 88.9%±2.4% | **94.0%±0.8%** |
| | T12 | 60.8%±2.6% | 82.4%±2.3% | 95.1%±2.0% | 95.7%±2.1% | 96.1%±1.1% | 95.6%±0.8% | **97.8%±1.3%** |
| | T13 | 45.3%±4.5% | 48.8%±10.4% | 94.3%±1.7% | 96.0%±0.6% | 93.4%±1.4% | 93.4%±1.6% | **96.7%±1.2%** |
| | Average | 45.5%±7.3% | 50.0%±10.7% | 89.2%±2.5% | 91.5%±2.3% | 94.3%±2.1% | 92.2%±2.2% | **96.3%±1.1%** |
| SNR=-5 | T01 | 40.3%±9.9% | 35.2%±4.3% | 56.4%±8.5% | 59.1%±2.8% | 74.5%±5.5% | 58.0%±6.1% | **86.1%±2.4%** |
| | T02 | 37.2%±4.0% | 38.0%±5.6% | 44.8%±6.4% | 57.5%±6.5% | 79.7%±2.0% | 79.5%±5.0% | **85.1%±1.9%** |
| | T03 | 16.8%±1.9% | 36.3%±3.5% | 49.3%±1.3% | 58.9%±5.3% | 62.4%±2.0% | 47.9%±2.9% | **80.8%±1.7%** |
| | T10 | 31.5%±3.1% | 35.3%±3.3% | 35.7%±2.2% | 49.1%±7.2% | 73.3%±2.4% | 35.1%±1.7% | **87.5%±1.8%** |
| | T12 | 31.9%±7.9% | 34.1%±3.8% | 81.4%±3.6% | 80.8%±1.4% | 79.1%±2.5% | 45.3%±4.1% | **82.7%±2.0%** |
| | T13 | 28.4%±5.4% | 38.9%±7.0% | 38.8%±5.4% | 55.6%±4.4% | 70.2%±3.9% | 46.7%±6.4% | **81.3%±0.9%** |
| | Average | 31.0%±5.4% | 36.3%±4.6% | 51.1%±4.6% | 60.2%±4.6% | 73.2%±3.1% | 52.1%±4.4% | **83.9%±1.8%** |

To visualize the extracted features and validate the domain alignment capability of different methods, we used the t-distributed stochastic neighbor embedding (t-SNE) method [47]. Fig. 10 illustrates the visualization results of the feature distribution on a randomly selected task T01 using various methods, with an SNR of -5 for the target domain data. For clarity, we only display the results from four methods. Different fault types are represented by different colors, while different domains are distinguished by different shapes. This transfer learning task is particularly challenging due to the presence of severe noise interference. As depicted in Fig. 10, conventional UDA methods exhibit significant domain misalignment, indicating a difficulty in aligning the feature distributions between the source and target domains due to the noise impact. In contrast, the SFDANN method demonstrates a remarkable ability to align the domains and perform accurate classification. The visualization confirms the strong domain alignment and classification capabilities of the SFDANN method in the face of challenging noise interference.

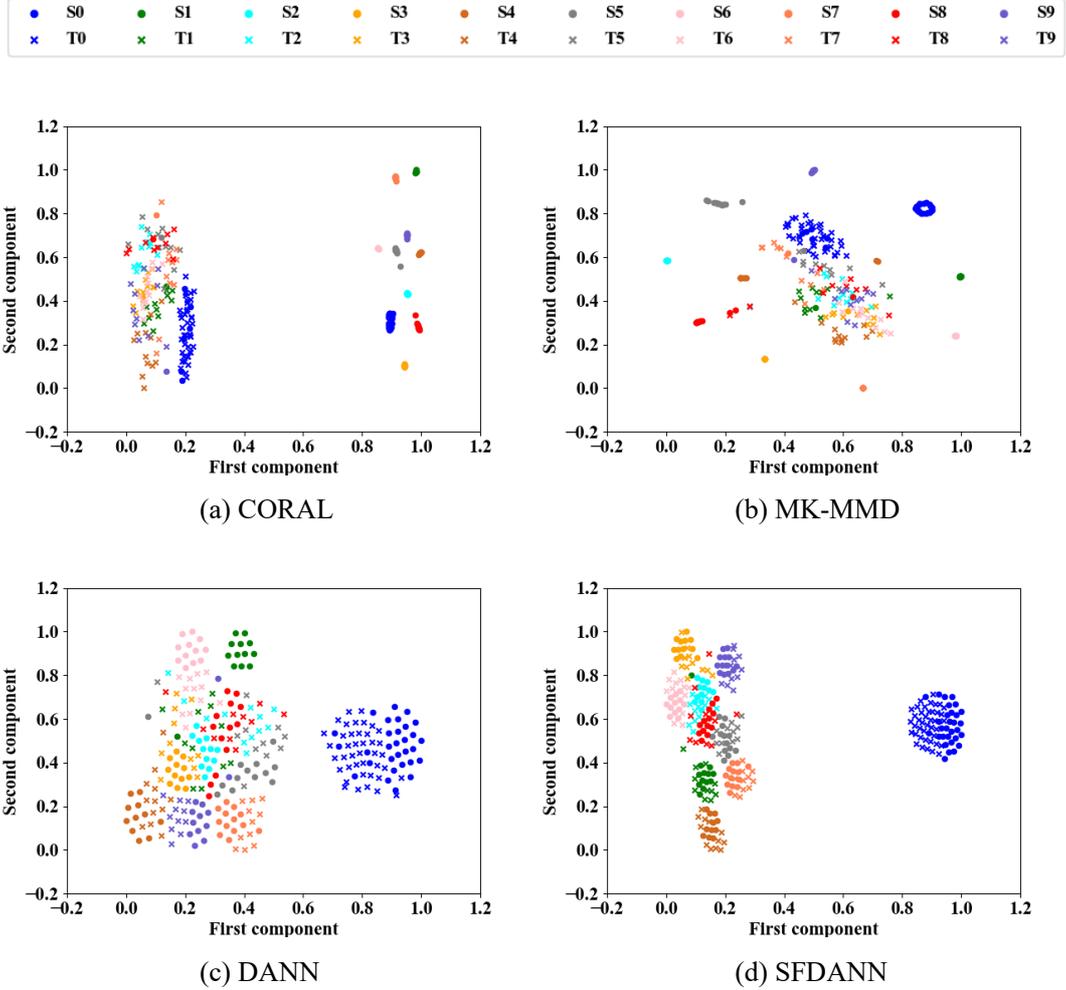

Fig. 10. Visualization of features using t-SNE for the bearing case with transfer task T01.

*(2) Diagnosis results of Case Study 2 (Slab track dataset)*

Table 3 presents the diagnosis results for Case Study 2, where the UDA methods are applied on transfer tasks from labeled numerical simulation data (source domain) to unlabeled field measurement data (target domain) at different train speeds. The proposed SFDANN method achieves the highest fault diagnosis accuracy across all transfer tasks. SFDANN demonstrates an average accuracy of 96.1% for all cross-domain tasks, surpassing the best performing method by 13.3% and the worst performing method by 26.3%. These results illustrate the exceptional fault diagnosis performance of the SFDANN method in transferring knowledge from simulation scenarios to real-world scenarios in complex industrial systems.

Table 3. Recognition accuracy of slab track health state.

| Speed | CNN | CORAL | MK-MMD | JAN | DANN | CDANN | SFDANN |
|---|---|---|---|---|---|---|---|
| 20 km/h | 60.0%±8.2% | 83.7%±9.3% | 79.3%±3.8% | 73.3%±3.6% | 63.0%±7.4% | 68.2%±3.0% | **100.0%±0** |
| 40 km/h | 71.1%±4.3% | 80.0%±10.1% | 77.8%±12.6% | 77.8%±8.4% | 72.6%±1.8% | 69.6%±3.6% | **96.3%±4.1%** |
| 60 km/h | 68.2%±3.0% | 81.5%±8.1% | 78.5%±6.4% | 70.4%±5.7% | 79.3%±6.0% | 80.0%±5.0% | **96.3%±4.1%** |
| 80 km/h | 80.0%±4.4% | 85.9%±3.6% | 84.4%±5.4% | 90.4%±1.8% | 85.2%±6.6% | 85.2%±5.2% | **91.9%±2.8%** |
| Average | 69.8%±5.0% | 82.8%±7.8% | 80.0%±7.1% | 78.0%±4.9% | 75.0%±5.5% | 75.8%±4.2% | **96.1%±2.8%** |

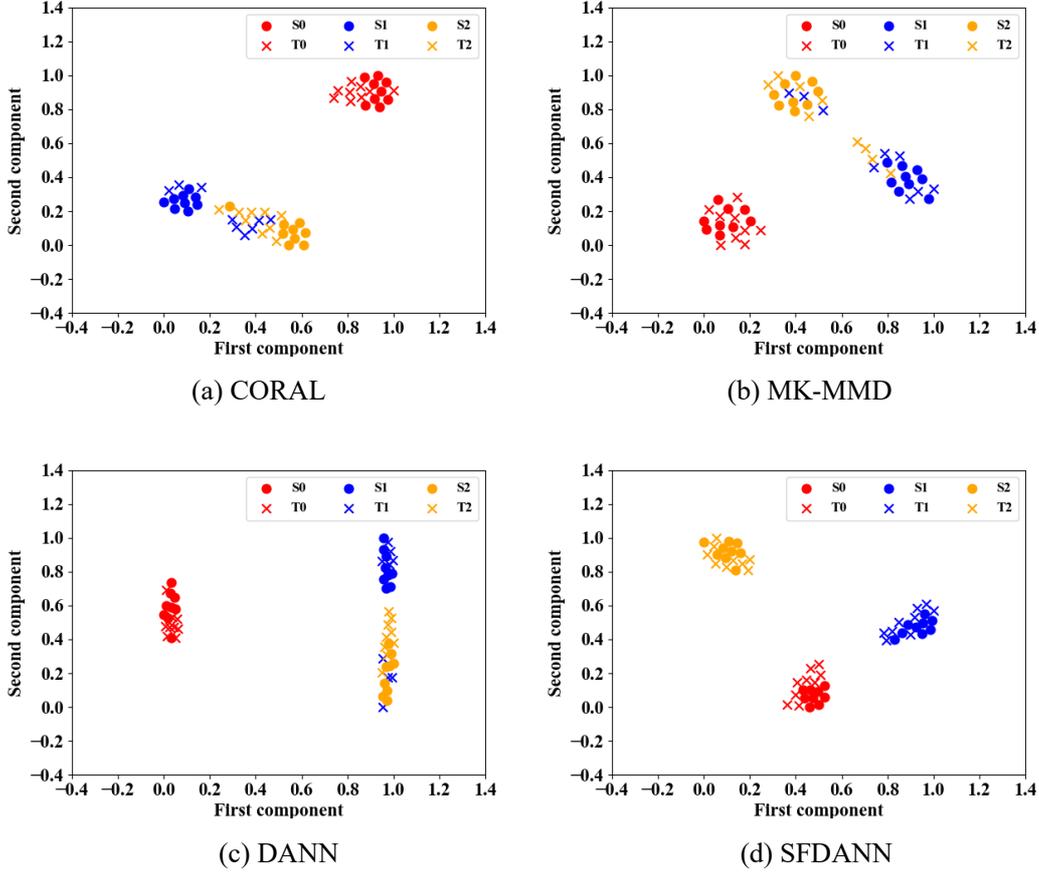

(a) CORAL  (b) MK-MMD  (c) DANN  (d) SFDANN

Fig. 11. Visualization of features using t-SNE for the slab track case at a train speed of 60km/h.

Similar to Case Study 1, we used the t-SNE method to evaluate the domain alignment capabilities of different methods by visualizing the extracted features. To ensure clarity, in Fig. 11, we only present the results for four methods, using the case of train speed 60km/h as an example. As observed from the figure, the conventional UDA methods fail to achieve complete domain alignment. However, the SFDANN method demonstrates robust domain alignment and classification capability.

*4.2.2. Ablation study*

To assess the influence of the LWPT and WPT modules on the performance of the SFDANN model, we conducted an ablation study in this section. In the Section 4.2.1, we already compared SFDANN with a variant that uses two WPT modules instead of the combination of LWPT and WPT modules, which corresponds to the classical DANN. In this section, we introduce another variant of SFDANN, referred to as SFDANN-v, where the combination of LWPT and WPT modules in SFDANN is replaced with two LWPT modules. Moreover, we added different levels of Gaussian white noise to source domain of CWRU dataset as a supplement of Case study 1, where the noisy environments is only considered for target domain. The performance of SFDANN with two distinct data input strategies is evaluated in this section to analyze the impact of data input strategy on the alignment of two domains.

Table 4. Fault diagnosis accuracy using CWRU dataset in ablation study.

| Task | | | SFDANN-v | SFDANN | |
|---|---|---|---|---|---|
| | | | | S→LWPT | T→LWPT |
| No noise | | T01 | 99.7%±0.3% | **99.9%±0.3%** | **99.9%±0.3%** |
| | | T02 | 99.7%±0.3% | **99.9%±0.3%** | 99.7%±0.3% |
| | | T03 | **99.9%±0.3%** | 99.7%±0.3% | 99.7%±0.3% |
| | | T10 | 99.7%±0.3% | **99.9%±0.3%** | **99.9%±0.3%** |
| | | T12 | 99.7%±0.3% | 99.7%±0.3% | **99.9%±0.3%** |
| | | T13 | 99.8%±0.3% | **99.9%±0.2%** | 99.9%±0.3% |
| | | Average | **99.8%±0.3%** | **99.8%±0.3%** | **99.8%±0.3%** |
| SNR=0 | Source | T01 | 98.1%±0.6% | **98.4%±0.5%** | 98.1%±1.1% |
| | | T02 | 98.7%±1.1% | **98.8%±1.0%** | 98.4%±0.8% |
| | | T03 | **99.5%±0.5%** | 99.2%±0.5% | 97.7%±1.9% |
| | | T10 | 96.2%±2.1% | **96.3%±0.9%** | 95.5%±0.9% |
| | | T12 | 97.6%±1.1% | 98.3%±0.9% | **98.6%±0.6%** |
| | | T13 | 99.5%±0.6% | **99.9%±0.3%** | 99.7%±0.3% |
| | Target | T01 | 94.4%±2.1% | 95.4%±1.0% | **96.1%±1.0%** |
| | | T02 | 95.8%±1.7% | 96.3%±1.1% | **97.1%±1.0%** |
| | | T03 | 93.4%±1.5% | 93.7%±2.5% | **96.2%±1.3%** |
| | | T10 | **94.4%±1.4%** | 93.8%±1.3% | 94.0%±0.8% |
| | | T12 | 96.9%±1.4% | 96.7%±0.5% | **97.8%±1.3%** |
| | | T13 | 95.8%±0.8% | 96.4%±2.2% | **96.7%±1.2%** |
| | | Average | 96.7%±1.2% | 96.9%±1.1% | **97.2%±1.0%** |
| SNR=-5 | Source | T01 | 91.0%±0.9% | **97.0%±1.0%** | 95.5%±0.7% |
| | | T02 | 87.6%±3.2% | **96.5%±0.7%** | 94.4%±1.8% |
| | | T03 | 91.1%±1.1% | **94.2%±1.9%** | 92.5%±0.7% |
| | | T10 | 88.8%±2.0% | **93.9%±1.1%** | 90.9%±1.5% |
| | | T12 | 92.2%±0.6% | **96.2%±1.5%** | 93.4%±1.4% |
| | | T13 | 92.5%±0.9% | **97.2%±1.1%** | 93.5%±2.7% |
| | Target | T01 | 79.6%±4.7% | 83.5%±3.0% | **86.1%±2.4%** |
| | | T02 | 74.7%±3.3% | 81.3%±1.0% | **85.1%±1.9%** |
| | | T03 | 72.1%±5.7% | 78.7%±1.0% | **80.8%±1.7%** |
| | | T10 | 83.2%±1.5% | 84.6%±1.5% | **87.5%±1.8%** |
| | | T12 | 78.2%±1.7% | 80.3%±1.7% | **82.7%±2.0%** |
| | | T13 | 77.1%±4.0% | 79.1%±3.2% | **81.3%±0.9%** |

|  | Average | 84.0%±2.5% | 88.5%±1.6% | **88.6%±1.6%** |

Note: "S→LWPT" indicates the data input strategy of feeding the source domain data into the LWPT module of smart filter; "T→LWPT" indicates the data input strategy of feeding the target domain data into the LWPT module of smart filter; "Source" means that the noise is added to the source domain; "Target" means that the noise is added to the target domain.

Table 5. Fault diagnosis accuracy using slab track dataset in ablation study.

| Speed | SFDANN-v | SFDANN | |
| --- | --- | --- | --- |
| | | S→LWPT | T→LWPT |
| 20 km/h | 87.4%±5.5% | 99.3%±1.5% | **100.0%±0** |
| 40 km/h | 75.6%±1.8% | 91.1%±6.5% | **96.3%±4.1%** |
| 60 km/h | 82.2%±1.5% | 91.9%±8.2% | **96.3%±4.1%** |
| 80 km/h | 86.7%±3.8% | 89.6%±1.5% | **91.9%±2.8%** |
| Average | 83.0%±3.2% | 93.0%±4.4% | **96.1%±2.8%** |

Note: "S→LWPT" indicates the data input strategy of feeding the source domain data into the LWPT module of smart filter; "T→LWPT" indicates the data input strategy of feeding the target domain data into the LWPT module of smart filter.

The experimental results of SFDANN-v and SFDANN on the bearing dataset and the slab track dataset are presented in Table 4 and 5. These results demonstrate that the overall performance of SFDANN is superior to that of SFDANN-v. On the bearing dataset, when there is no Gaussian white noise in the data, both SFDANN and SFDANN-v exhibit nearly identical fault diagnosis accuracy. However, in the presence of Gaussian white noise in either the source domain or the target domain, SFDANN consistently outperforms SFDANN-v in most cases. For the slab track dataset, SFDANN demonstrates better domain adaptation ability compared to SFDANN-v. This suggests that preserving either the source domain data or the target domain data, and promoting the other domain data to resemble the preserved data, yields improved domain alignment results compared to using two LWPT modules to align both domains simultaneously. This observation can be attributed to the fact that the combination of LWPT and WPT modules has fewer learnable parameters compared to two LWPT modules. As a result, it reduces training difficulty and enhances training stability, leading to better performance in SFDANN.

Table 4 additionally exhibits the fault diagnosis results of bearings using two data input strategies with the SFDANN method. It can be observed that the diagnostic accuracy is similar when there is no Gaussian noise in the dataset, regardless of the input strategy employed. However, when noise is present in the source domain, the overall diagnostic accuracy is higher when the noisy source domain data is inputted into the LWPT module of the smart filter, while the noise-free target domain data is inputted into the WPT module. Conversely, when noise exists in the target domain, the diagnostic performance is generally better when the noisy target domain data is inputted into the LWPT module. This phenomenon can be attributed to the guidance loss, which drives the time-frequency characteristics of the input data in the LWPT module to closely align with those of the input data in the WPT module. If the data inputted to the WPT module is less affected by noise, the data inputted to the LWPT module will undergo a certain level of denoising during the model training process, leading to improved fault diagnosis performance compared to conventional UDA methods. However, if the data inputted to the WPT module is heavily impacted by noise, the data inputted to the LWPT module experiences a certain level of noise augmentation during training. While the SFDANN with this data input strategy still outperforms conventional UDA methods in most cases, the degree of performance improvement is not as significant as that achieved by the data input strategy capable of denoising the noisy data.

Similar phenomenon can be observed in the state recognition results of slab tracks. Table 5 indicates that the state recognition accuracy obtained by inputting the field measurement data into the LWPT module of the smart filter and the numerical simulation data into the WPT module is higher than that obtained by the other data input strategy. This is mainly because numerical simulation signals are purer and easier to be classified correctly by the classifier than field measurement signals. Therefore, promoting the field measurement signals similar to the numerical simulation signals is more conducive to fault diagnosis than promoting the numerical simulation signals similar to the field measurement signals. As a result, it can be concluded from Table 4 and 5 that the optimal data input strategy for SFDANN is to feed the data from the noisier domain into the LWPT module of the smart filter, while inputting the data from another domain into the WPT module of the smart filter.

*4.2.3. Effect of smart filter*

To further investigate the effectiveness of the smart filter in SFDANN, we analyzed the spectral changes of data passing through the smart filter. We focused a UDA task from operation condition 0 to 2 in the bearing case study, where the Gaussian white noise was added to the target domain data to create noisy signals with an SNR of -5. In this analysis, we input noisy target domain data (i.e., data from operation condition 2) into the LWPT module, while the source domain data (i.e., data from operation condition 0) was input into the WPT module. After training the SFDANN model, we selected the original source domain data, original target domain data, reconstructed source domain data, and reconstructed target domain data corresponding to label 2 as examples to explore the effect of the smart filter. To assess the impact of the smart filter, we calculated the mean frequency spectra for these four groups of data and plotted them in Fig. 12(a), while the corresponding one-third octave spectra were shown in Fig. 12(b).

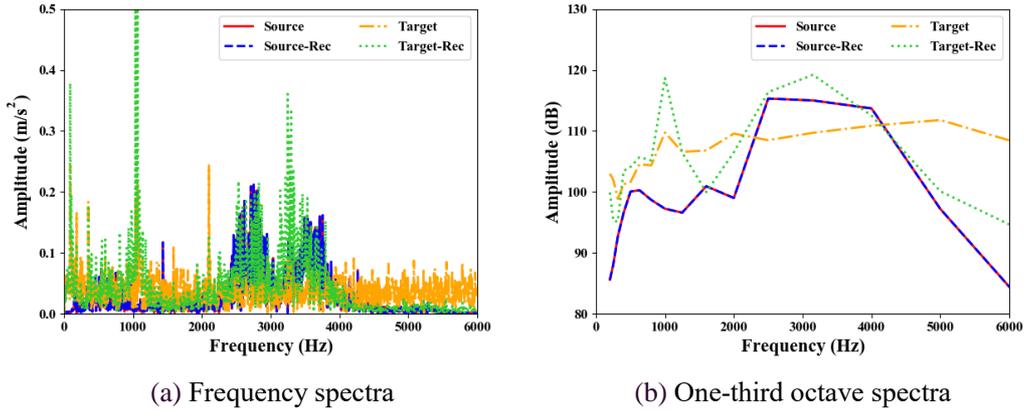

(a) Frequency spectra  (b) One-third octave spectra

Fig. 12. Source and target domain data corresponding to label 2 from CWRU dataset: "Rec" denotes "Reconstructed".

The frequency spectra of the original and reconstructed source domain data completely overlap in Fig. 12, indicating that there is no change in the source domain data before and after passing through the smart filter. However, the original target domain data is severely distorted by noise, leading to a significant deviation in its frequency spectrum compared to that of the source domain data. The reconstructed target domain data exhibits a frequency spectrum shape that is closer to that of the source domain data when compared to the original target domain data. This suggests that the target domain data has increased similarity with the source domain data after passing through the smart filter. Such alignment between the source and target domains is beneficial for subsequent feature extractors to capture shared features and further improve fault diagnosis accuracy.

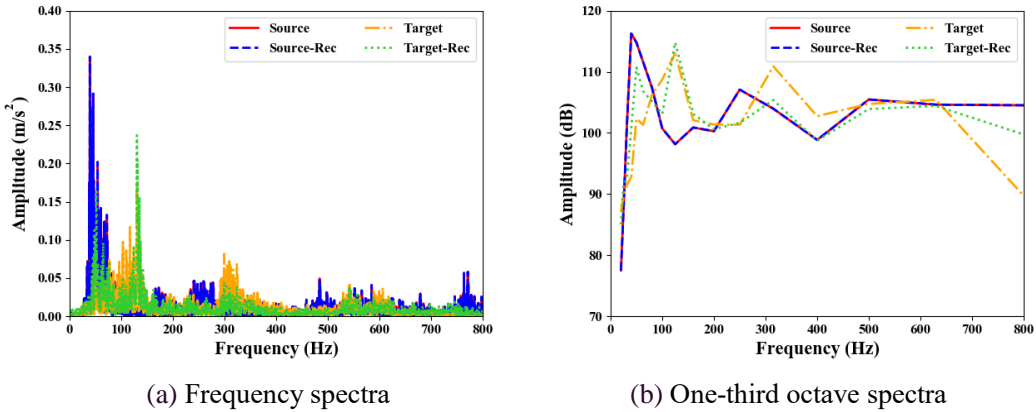

(a) Frequency spectra  (b) One-third octave spectra

Fig. 13. Source and target domain data corresponding to label 2 from slab track dataset: "Rec" denotes "Reconstructed".

For the slab track case, we select a UDA task with a train speed of 60 km/h as an example. In this case, the measurement data serves as the target domain data and is input into the LWPT module, while the numerical simulation data serves as the source domain data and is input into the WPT module. After the model training, we obtain original and reconstructed data with label 2 for both domains. Their mean frequency spectra and corresponding one-third octave spectra are plotted in Fig. 13. From Fig. 13(b), it is evident that, at the one-third octave scale, the shape of the reconstructed source domain data is closer to that of the target domain data compared to the original source data. This demonstrates the effectiveness of the smart filter in promoting the similarity between the data from the two domains. This increased similarity is beneficial for SFDANN model to learn domain-invariant and discriminative features, leading to improved performance in fault diagnosis tasks.

## 5. Conclusion

We propose SFDANN, an unsupervised domain adaptation method to address the challenge of domain misalignment in fault diagnosis scenarios characterized by different noise levels. SFDANN consists of a smart filter based on LWPT and WPT, a feature extractor, a domain discriminator, and a classifier. The effectiveness of SFDANN is demonstrated through case studies on bearing fault diagnosis in noisy environments and on simulation to real transfer of slab track fault diagnosis in a train-track-bridge coupling vibration system. The latter case study involves the transfer of models from labeled numerical simulations to unlabeled field measurements. Furthermore, this study analyzes the impact of two data input strategies, (1) inputting source and target domain data separately into LWPT and WPT modules, and (2) the reverse input strategy, on fault diagnosis results.

The results indicate that the designed smart filter can dynamically approximate the source and target signals in the time-frequency domain, allowing SFDANN to learn domain-invariant and discriminative features. Inputting the data from the domain with more severe noise interference into the LWPT module of the smart filter, and the data from the other domain into the WPT module, proves more beneficial for improving the overall final fault diagnosis accuracy. Furthermore, the proposed SFDANN method outperforms other UDA methods in terms of domain adaptability and diagnostic performance in industrial scenarios with unknown types and levels of noise.

In comparison to other UDA methods, the primary drawback of our proposed method is the additional computational cost incurred during the training of the newly introduced module, smart filter. However, it is important to note that the increase in the computational cost is only significant when the number of signal decomposition layers becomes excessively large. Our future research will focus on partial domain adaptation for fault diagnosis. Additionally, we plan to evaluate the performance of SFDANN on other challenging fault knowledge transfer tasks, such as utilizing signals measured by different types of sensors. The proposed SFDANN algorithm is very versatile and flexible which makes it applicable across a wide range of potential applications, especially those susceptible to noise interference. This applicability extends beyond the specific fault diagnosis scenarios showcased in this study. Illustrative examples include transfer learning tasks in noisy speech recognition and health-oriented applications, particularly involving the processing of electroencephalogram (EEG) and electrocardiogram (ECG) data.

## Acknowledgement

This study was supported by the National Natural Science Foundation of China (grant number 52178432) and China Scholarship Council (grant number 202106260178).